# Self-Consistent Hopping Theory of Activated Relaxation and Diffusion of Dilute Penetrants in Dense Crosslinked Polymer Networks

Baicheng Mei [a,d], Tsai-Wei Lin [c,d], Charles E. Sing [a,c,d] and Kenneth S. Schweizer [*a,b,c,d]

[a] Department of Materials Science, University of Illinois, Urbana, IL 61801
[b.] Department of Chemistry, University of Illinois, Urbana, IL 61801
[c.] Department of Chemical & Biomolecular Engineering, University of Illinois, Urbana, IL 61801
[d.] Materials Research Laboratory, University of Illinois, Urbana, IL 61801

[*] *kschweiz@illinois.edu*

**Abstract**

We generalize and apply a microscopic force-level statistical mechanical theory of the activated dynamics of dilute spherical penetrants in glass-forming liquids to study the influence of permanent crosslinking in polymer networks on the penetrant relaxation time and diffusivity over a wide range of temperature and crosslink density. The theory treats network crosslinkers as locally pinned or vibrating sites. Calculations are performed for model parameters relevant to recent experimental studies of an nm-sized organic molecule diffusing in crosslinked poly-(n-butyl methacrylate) networks. The theory predicts the penetrant alpha relaxation time increases exponentially with the crosslink fraction ($f_n$) dependent glass transition temperature, $T_g$, which grows roughly linearly with the square root of crosslink density. Moreover, $T_g$ is also found to be proportional to a geometric confinement parameter defined as the ratio of the penetrant diameter to the mean network mesh size. The decoupling ratio of the penetrant to polymer Kuhn segment alpha relaxation times displays a complex non-monotonic dependence on crosslink density and temperature that can be well collapsed based on the variable $T_g(f_n)/T$. The microscopic mechanism for activated penetrant relaxation is elucidated based on the theoretically predicted crosslink fraction dependent coupled local cage and nonlocal collective elastic barriers. A model for the penetrant diffusion constant that combines activated segmental dynamics and entropic mesh confinement is proposed which results in a significantly stronger suppression of mass transport with degree of effective supercooling than predicted for the penetrant alpha time. This behavior corresponds to a new polymer network-based type of "decoupling" of diffusion and relaxation. In contrast to the diffusion of larger nanoparticles in high temperature rubbery networks, our analysis in the deeply supercooled regime suggests that for the penetrants studied the mesh confinement effects are of secondary importance relative to the consequences of crosslink-induced slowing down of glassy activated relaxation. Support for the theoretical predictions from our complementary simulation study is briefly discussed.

## I. Introduction

Understanding and controlling activated molecular ("penetrant") transport in dense and cold polymer melts, crosslinked networks, and glasses is a problem of high basic polymer physics interest[1-14] that also underpins advanced separation membrane applications[10-27]. However, even in the *dilute* penetrant limit of interest in this work, existing theoretical models are highly phenomenological, often built on the difficult to uniquely define and unambiguously predict concept of "free volume".[13, 27-33] Recently, two of us have pursued the construction of a force-level, microscopic statistical mechanical theory (the Self Consistent Cooperative Hopping (SCCH) theory) for penetrant activated dynamics in polymer melts [34]. The foundation is the Elastically Collective Nonlinear Langevin Equation (ECNLE) theory[35, 36] for the thermally activated structural (alpha) relaxation process on the Kuhn segment scale in one-component polymer liquids[37]. This approach is microscopic in the sense that it causally relates pair interactions, thermodynamic state, and packing structure as embedded in a dynamic free energy which quantifies kinetic constraints. The alpha relaxation is a coupled local-nonlocal activated process characterized by a local cage barrier ($F_{B,K}$) associated with large-amplitude hopping coupled with collective longer-range (nonlocal) small displacements of all segments outside the cage characterized by an elastic barrier ($F_{el,K}$). Recent studies support the predictions of this approach for a large number of real-world polymeric liquids, including olefins, dienes, siloxanes, and vinyl polymer chemistries.[38-40] Chemical crosslinking introduces new kinetic constraints and additional dynamical slowing down. ECNLE theory has been very recently extended to polymer networks, and has provided an excellent understanding of both experiments and coarse grained molecular dynamics simulations on poly(n-butylacrylate) (PnBA) networks over a wide range of crosslink densities and temperatures.[41]

SCCH theory for dilute hard sphere penetrants (diameter, $d$) in a hard-sphere matrix [42, 43] or semi-flexible polymer melts [34] under structurally equilibrated conditions has been extensively applied, and many of its predictions agree well with simulations [7, 42] and experiments [44]. The penetrant activation barrier and mean hopping time, and the extent of coupling of penetrant transport with the early, medium, and late stages of the matrix structural relaxation process, are self-consistently predicted based on two *coupled* dynamic free energies.[34, 42, 43, 45]

The goal of the present work is to formulate (and apply) SCCH theory to treat permanent crosslinked polymer networks. We are especially motivated by recent measurements of Evans and coworkers on the diffusion of a dilute aromatic molecule [N,N'-Bis(2,5-di-tert-butylphenyl)-3,4,9,10-perylenedicarboximide (BTBP)] in crosslinked PnBA networks over a wide range of crosslink densities at fixed temperature.[9] The experiments found that the penetrant alpha time (as deduced from the dielectric loss response) and diffusion constant both vary exponentially with the crosslink fraction ($f_n$) dependent network glass transition temperature $T_g(f_n)$, but their quantitative dependences on $T_g(f_n)$ and $T$ are different, a form of "decoupling" in the language of glass physics. The precise physical mechanisms at play remain unclear. For example, is the apparent decoupling a consequence of entropic mesh confinement constraints on penetrant motion in crosslinked networks which modifies the long-time mass transport process more than the influence of polymer segmental alpha relaxation? Alternatively, is the glass physics mechanism associated with the dynamic heterogeneity (DH)[46-48] driven breakdown of the Stokes-Einstein breakdown relation [49-54] in supercooled liquids relevant?

The mesh confinement effect has been studied theoretically[55], experimentally[9], and with simulation[56, 57] for the problem of the motion of a dilute spherical nanoparticle (larger than typical molecules) in high temperature rubbery polymer networks. A key variable is the confinement parameter, $C = d/a_x$, where $a_x$ is the mean mesh size that decreases with crosslink density. For example, the polymer physics model of Cai-Panykov-Rubinstein (CPR) [55] predicts that when mesh confinement is strong enough ($C>1$) the nanoparticle diffusion constant is:

$$D_p \propto \frac{d^2}{\tau_R} \exp(-C^2)/C \equiv \frac{d^2}{\tau_R} X \tag{1}$$

where $\tau_R \sim N_x^2 \tau_\alpha$ is the Rouse time of a network strand composed of $N_x$ segments, the segmental alpha time, $\tau_\alpha$, depends on temperature but is assumed to *not* vary with crosslink density, and the quantity $X(C)$ (defined above) quantifies the explicit entropic effect of a physical mesh. Such a model is most relevant to the high temperature rubbery polymer network regime. Recently, a simulation study of spherical penetrant dynamics in high temperature semi-dilute polymer networks has been performed [57], and found $\log(1/CD_p) \propto C^2$ for *both* $C > 1$ and $C < 1$. By revisiting their data, we find that the prefactor between $\log(1/CD_p)$ and $C^2$ is significantly larger than $\log(e)$ in the CPR model, suggesting the kinetic factor $\frac{d^2}{\tau_R}$ in Eq.(1) must also contribute significantly to the crosslink density dependence of the penetrant diffusion constant and appears to also vary exponentially with $-C^2$. Qualitatively, this is expected within the SCCH theory framework since the segmental alpha time in polymer networks increases strongly with crosslink fraction (as observed in experiment and simulation[41]) which must significantly impact the penetrant alpha relaxation process.

In this article we propose and test the idea that the penetrant diffusion constant in Eq.(1) should scale inversely with the activated penetrant alpha time as:

$$D_p \propto \frac{d^2}{\tau_{\alpha,p}} \exp(-C^2)/C \equiv \frac{d^2}{\tau_{\alpha,p}} X \propto \exp(-BC^2)/C \tag{2}$$

where $\tau_{\alpha,p}$ is both temperature and crosslink density dependent, and $B$ is a chemistry specific numerical factor. Since in most real world situations the molecular penetrants do not obey the strong mesh confinement inequality $C>1$, other workers [56] have considered, and explored with simulation, an alternative model where the penetrant entropic barrier scales linearly with the confinement parameter, $D_\mathrm{p} \propto \frac{d^2}{\tau_\mathrm{R}}\exp(-b'C)$, where $b'$ is a system specific constant. If one adopts this description of mesh confinement effects, then Eq.(2) becomes:

$$D_\mathrm{p} \propto \frac{d^2}{\tau_{\alpha,\mathrm{p}}}\exp(-b'C) \propto \exp(-EC) \qquad (3)$$

Here, and in our companion simulation paper [58], we adopt SCCH theory to quantitatively analyze the effect of crosslinking on the penetrant alpha relaxation time and diffusion constant in the context of the PnBA networks[41]. In general, the penetrant alpha time and diffusion constant are high-dimensional functions, depending on variables such as temperature, penetrant chemistry (size and shape), penetrant-polymer interaction, polymer persistence or Kuhn length, and degree of crosslinking (or mesh size). Here we consider a single spherical penetrant of diameter that mimics the nonpolar organic dye molecule BTBP studied experimentally [9, 59] by choosing its diameter to be $d=1.12$nm which reproduces the space filling volume of BTBP of 735.88 Å$^3$. A purely repulsive penetrant-polymer interaction and a semiflexible polymer chain model calibrated for PnBA networks are adopted. Our focus is on how temperature, crosslink density, and geometric mesh size impact the penetrant $\tau_{\alpha,\mathrm{p}}$ and $D_\mathrm{p}$. This choice of penetrant size and the range of crosslink fractions studied allows us to probe confinement parameters from well below to modestly above unity, as germane to the typical molecular penetrant problem.

The remainder of the article is structured as follows. In Section II we briefly describe ECNLE theory and its extension to crosslinked polymer networks, and present new calculations relevant to the penetrant work in this article. SCCH theory for liquid matrices is briefly reviewed in section III, extended to crosslinked networks, and foundational theoretical quantities required to predict penetrant alpha relaxation time and diffusion constant are presented. The penetrant relaxation time and its relationship to the network glass transition temperature and mesh confinement parameter are systematically studied in section IV. The question of the degree of decoupling between the penetrant and polymer segmental relaxation times is analyzed in section V. The role of crosslinking on the penetrant diffusion constant is studied in section VI, and the effects of crosslink-induced mesh confinement versus prolongation of network segmental relaxation on penetrant transports are disentangled. The article concludes in section VII with a summary and future outlook. Throughout the paper we comment on the comparison of our theoretical results with our companion simulation study [58] at the level of qualitative and semi-quantitative trends. Our intent is not to quantitatively test the theory with simulation since the polymer and penetrant models, and range of temperatures explored, are not the same. Rather, the two articles are meant to be complementary studies of common scientific problems.

We emphasize that nearly *all* the theoretical technical details, equations, derivations, numerical implementations, and discussion of limitations of the dynamic theories (ECNLE[35-37, 41] and SCCH[34, 42, 43, 45]) at the Kuhn scale, and the Polymer Reference Interaction Site Model (PRISM) integral equation theory[34, 37, 41] for the required structural pair correlation functions, have been thoroughly documented in the literature and are not repeated here. The present paper is the first to present the detailed extension of SCCH theory to crosslinked polymer networks, but its analog for polymer melts has been discussed previously in detail [34]. Thus, our review of ECNLE theory in section II and SCCH theory in section III focuses on their physical content and key parameters, with mathematical development presented only for the new aspects of SCCH theory required to treat crosslinked networks.

**II. ECNLE Theory of Structural Relaxation in Crosslinked Polymer Networks**

    **A. Theoretical Background for Homopolymer Melts and Crosslinked Networks**

The foundational starting point for predicting the activated dynamics of one-component spherical particle fluids is the single particle scalar displacement ($r$) - dependent dynamic free energy, $F_{\text{dyn}}(r)$. The negative gradient of this quantity determines an effective force in a stochastic nonlinear Langevin equation that controls the spatially local aspect of activated barrier hopping trajectories and a local cage barrier (right panel of Fig.1).[60] At high (low) enough packing fraction (temperature), the particle jump distance is predicted to be sufficiently large (denoted as $\Delta r_K$ in the Kuhn segment level implementation; see Fig.1) that in order for it to occur all the particles outside the cage must elastically displace in a dilational (cage expansion) collective manner by a small amount. This cage expansion sets the amplitude of the collective elastic displacement field [35, 36] and costs elastic energy. The resultant physical picture is that the alpha relaxation is a coupled local-nonlocal in space process (see middle panel of Fig.1), characterized by local cage and longer-range collective elastic barriers.

Recently, we extended ECNLE theory to homopolymer melts by including the consequences of local chain connectivity and backbone stiffness on packing structure and dynamics [37, 41, 61]. Motivated by the theoretical arguments of Zhou and two of us [37], the Kuhn segment is adopted as the relevant dynamical length scale for the alpha process, and thus the scalar displacement of the center-of-mass of rigid Kuhn segment (consisting of several bonded elementary sites) from its initial position, $r_K(t)$, is the central dynamical variable. The Kuhn segment is characterized by an effective *dynamically* correlated number of interaction sites or beads, $N_K$ (see left panel of Fig.1) in a semiflexible Koyama model[62, 63]. For typical experimental homopolymers, $N_K$ ~2-3, and in the present work we set $N_K = 2l_p$ with $l_p$ the persistence length. The sensibility of this model construction has been quantitatively tested in prior combined simulation-theory work [37].

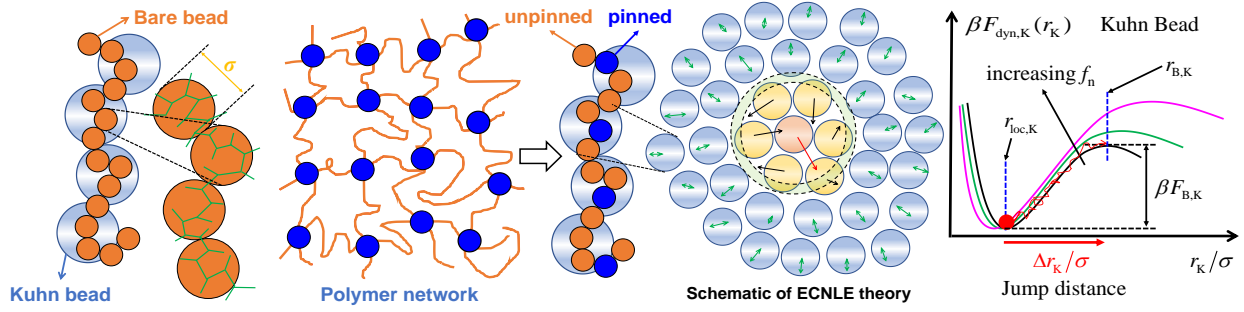

Fig.1. Left: Schematic of coarse-graining a polymer melt to the Kuhn segment scale and introduction of crosslinks based on pinning the elementary interaction sites or beads in a regular manner along the chain. Middle: Schematic of the physical ideas of the relaxation mechanism in ECNLE theory shown for simplicity for a spherical particle liquid: activated hopping involves a local cage barrier and a longer-range cooperative elastic displacement of all particles outside the cage which induces an additional nonlocal elastic barrier. Right: Schematic of key features of the dynamic free energy as a function of dimensionless displacement of a Kuhn segment and its evolution with increasing crosslink fraction, $f_n$. The localization length $r_{loc,K}$ decreases with $f_n$, while the local cage barrier, barrier location $r_{B,K}$, and jump distance $\Delta r_K = r_{B,K} - r_{loc,K}$ all increase.

The local cage barrier $F_{B,K}$ is obtained directly from the dynamic free energy which is constructed solely from knowledge of the inter- and intra- molecular equilibrium pair correlation functions. The elastic barrier has been derived to be: $\beta F_{el,K} \approx 2\pi r_{cage,K}^3 \Delta r_{eff,K}^2 \rho_K K_{0,K}$, where $K_{0,K}$ is the harmonic curvature (spring constant) at the localization length and the angularly averaged effective cage dilation amplitude is $\Delta r_{eff,K} \approx 3\Delta r_K^2/32 r_{cage,K}$; both of these quantities are computed based on the dynamic free energy. The Kuhn scale cage radius is $r_{cage,K} = N_K^{1/3} r_{cage,m}$ with $r_{cage,m}$ the location of the first minimum of polymer radial distribution function $g_{mm}(r)$ at the interaction site level and $\rho_K = \rho/N_K$ the number density of Kuhn segments.

Chemical crosslinking is modeled as follows [41]. First, it is assumed to not change the site-site intramolecular and intermolecular structural pair correlations. This simplification is consistent with a recent simulation study [64] that found the dimensionless compressibility of crosslink networks remains essentially unchanged compared to the melt. Second, a "neutral confinement" model of partial random pinning of particles (as widely studied in the glassy dynamics area [41, 61, [65]) is adopted which in the polymer network context mimics chemical crosslinking as a fraction ($f_\text{n}$) of dynamically immobile segments distributed in a regular manner along a chain, per the schematic in Fig.1. The ratio $n_\text{crosslink}/(n_\text{crosslink} + n_\text{monomer})$ defines the crosslink fraction $f_\text{n}$, where $n_\text{crosslink}$ and $n_\text{monomer}$ are the interaction site (bare bead) level number of crosslinked (pinned) and normal mobile (unpinned) beads, respectively. Determination of the corresponding dynamic partial structure factors which enter ECNLE theory as Debye-Waller factors [41, 61, 66] in the construction of the dynamic free energy is identical to prior work [41]. Crosslinkers modeled as pinned sites modify *all* aspects of ECNLE theory via the dynamic free energy.

The mean barrier hopping time $\tau_{\text{hop},K}$ is then computed based on the Kramers mean first passage time theory as [41, 42, 67, 68]

$$\tau_{\text{hop},K} = \frac{2\tau_{s,K} e^{\beta F_{\text{el},K}}}{\sigma_K^2} \int_{r_{\text{loc},K}}^{r_{\text{loc},K}+\Delta r_K} dr_K e^{\beta F_{\text{dyn},K}(r_K)} \int_{r_{\text{loc},K}}^{r_K} dr'_K e^{-\beta F_{\text{dyn},K}(r'_K)} \quad (4)$$

where $\tau_{s,K} = \sigma_K^2/D_{s,K}$ is the characteristic time scale of the very short time/length scale non-activated dynamical process that is unaffected by crosslinking for a Kuhn unit ($\sigma_K = N_K^{1/3}\sigma$) and the formula for $D_{s,K}$ is given elsewhere [35, 41, 42]. Using the PRISM integral equation theory of structural pair correlation functions as input to the dynamic free energy and Eq.(4), the mean Kuhn segment alpha time is computed numerically as $\tau_{\alpha,K} = \tau_{s,K} + \tau_{\text{hop},K}$. To compare our theoretical results with experimental or simulation results, we adopt the elementary timescale of a Newtonian liquid, $\tau_0 = 16\phi\sigma(\beta M/\pi)^{1/2}$ as the time unit, which is typically ~ 1 ps, per prior studies [35, 41].

**B. Mapping to Thermal Polymer Melts and Networks**

To determine the required equilibrium site-site pair correlation functions, PRISM theory with the Percus-Yevick (PY) closure is employed.[41, 61] The penetrant and polymer interaction sites are modeled as hard spheres. The intramolecular structure factor of an ideal semiflexible chain $\omega_{\mathrm{mm}}(q)$, penetrant-to-matrix size ratio $d/\sigma$, and polymer site reduced density or packing fraction enter as inputs to PRISM theory, and $\omega_{\mathrm{mm}}(q)$ is determined using the semiflexible discrete Koyama model[41, 62, 68]. This single chain statistical model is characterized by the bond length $l_{\mathrm{b}}$, bead diameter $l_{\mathrm{d}}$, and persistence length $l_{\mathrm{p}}$; we adopt its "tangent" version corresponding to $l_{\mathrm{b}} = l_{\mathrm{d}} = \sigma$. In our recent work on pure polymer network relaxation, a generic flexible chain persistence length of $l_{\mathrm{p}} = \frac{4}{3}\sigma$ was adopted [41]. In the present work, for the PnBA network the Kuhn length and monomer effective diameter are evaluated from experimental information [41] to be $l_{\mathrm{K}} = 1.72$nm and $l_{\mathrm{b}} = 0.72$nm, respectively, corresponding to a local aspect ratio $l_{\mathrm{K}}/l_{\mathrm{b}} = 2.4$, or in terms of the persistence length $l_{\mathrm{p}}/\sigma = 1.2$. Since in the tangent Koyama model one has $\frac{l_{\mathrm{K}}}{\sigma} = \frac{2l_{\mathrm{p}}}{\sigma} - 1$, the adopted mapping yields $\sigma = 1.23$nm, which is slightly larger than $\sigma = 1.03$nm if $l_{\mathrm{p}} = \frac{4}{3}\sigma$ is chosen. The core theoretical predictions for the $l_{\mathrm{p}}/\sigma = 1.2$ and $4/3$ models are very similar (see Appendix). In the main text we adopt the $l_{\mathrm{p}}/\sigma = 1.2$ model.

To address experimental systems under 1 atm isobaric conditions where temperature is the control variable, we adopt the well tested mapping approach that relates in a chemistry specific manner the polymer packing fraction to temperature.[39-41] Specifically, the dimensionless compressibility, $S_0(T)$, computed from PRISM theory for the semiflexible Koyama chain model[41, 62, 68] is equated to that obtained from experimental equation of state (EOS) data[44, 69] for PnBA melts.[39-41] The result of this mapping is shown in the inset of Fig.2. A nearly linear dependence is valid from the very high temperature regime (>400K) through the modestly supercooled regime, down to the glass transition temperature. The main frame of Fig.2 shows $S_0$ can be accurately represented by the simple analytic formula $S_0^{-1} = N_{\mathrm{s}}(B/T - A')^2$ which has been derived from the van der Waals model[70] EOS. As discussed previously[38, 70, 71], reasonable variation of $N_s$ does not affect our analysis since the crucial aspect is the $T$-dependence of $S_0$ that is determined by the EOS parameters $A'$ and $B$, and $N_s$ enters only as a *T-independent* prefactor. The final step of our model construction is to set $N_s = 7.75$ to reproduce the experimental glass transition temperature of $T_g = -45$ °C of pure PnBA melts [9].

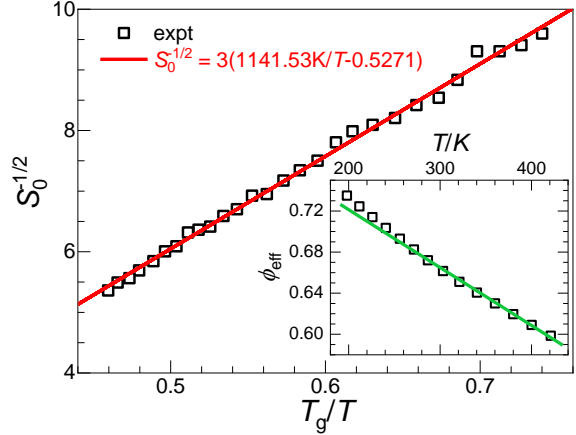

Fig. 2. Main: Experimental dimensionless compressibility data of a PnBA melt at 1 atm pressure vs scaled inverse temperature in a representation suggested by the van der Waals model [70]. One sees the data agrees very well with expectations that $S_0^{-1/2}$ is a linear function of $T_g/T$ in the wide range $T_g/T$=0.46-0.74. The red line corresponds to the analytic representation [70] $S_0^{-1} = 9(1141.53K/T - 0.5271)^2$, which is employed in the main text to map the hard chain model to its thermal PnBA analog. Inset: Relationship between effective packing fraction from PRISM theory and temperature for PnBA deduced from the mapping procedure [39-41].

### C. Dynamic Relaxation of Kuhn Segments and Glass Transition Temperature.

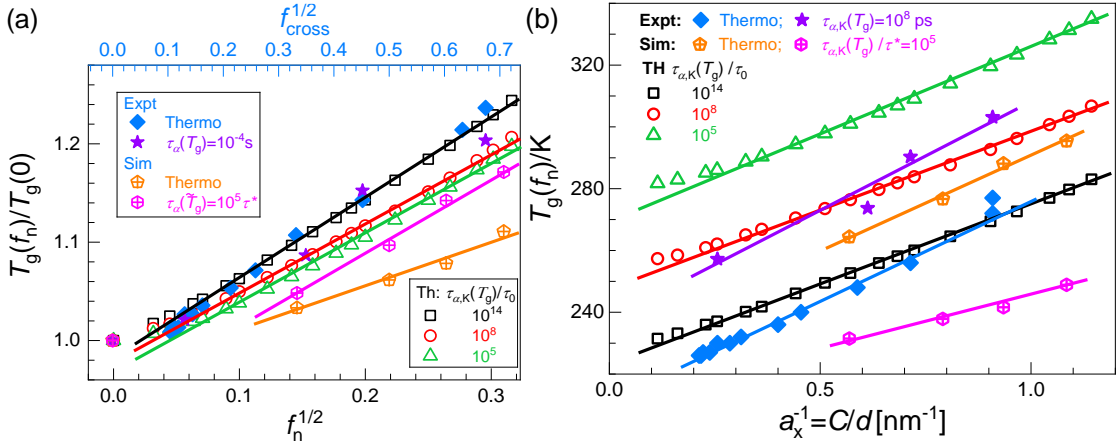

Fig.3. (a) Comparisons between scaled glass transition temperatures from experiment, simulation, and theory ($\frac{l_p}{\sigma} = 1.2$ model) extracted and defined in two ways as indicated (dynamic and thermodynamic), $T_g(f_n)/T_g(0)$, plotted as a function of the square root of crosslink fraction. The experimental and simulated crosslink fractions are taken to be exactly the same and the theory analog obeys $f_n = 0.19 f_{cross}$ [41]. (b) Glass transition temperature $T_g(f_n)$ in Kelvin as a function of inverse mesh size. Experimental and simulation mesh sizes are directly measured where the bond length $l_b = 0.72$ nm was used in simulation as the length unit [41] (based on assigning a monomer volume as $V_{mol} = \pi\sigma^3/6$ in contrast to the prior work that adopted $V_{mol} = \sigma^3$) while in the theory analysis the mesh size is calculated using $a_x^{-1} = A(f_n/0.19)^{1/2}/\sigma$ with $A$=1.94 (see main text).

We first consider how the degree of crosslinking affects the glass transition temperature $T_g(f_n)$ defined in the theory as when Kuhn segment alpha time reaches a fixed value relevant to a specific experiment or simulation; here we employ $\tau_{\alpha,K}(T_g)/\tau_0 = 10^{14}, 10^8$, and $10^5$ to illustrate our predictions based on a vitrification criterion in the laboratory deeply supercooled, intermediate supercooled, and weakly supercooled (per simulations) regimes, respectively. This issue was studied in our previous publication for the $l_p/\sigma = 4/3$ Koyama model where we found $T_g(f_n)$ grows non-linearly with $f_n$, but no precise functional form identified.[41, 72] Here, as shown in Fig.3a (and also its analogue for the $l_p/\sigma = 4/3$ Koyama model, not shown), we report a *new* discovery of a square-root relationship, $T_g(f_n) \propto f_n^{1/2}$, with only small deviations evident in the very low crosslink density regime. Adoption of different $T_g$ criteria do not change the robustness of this relation. Fig.3a also shows that the previously obtained experimental and simulation $T_g(f_{cross})$ data[41] obey well the relation $T_g(f_{cross}) \propto f_{cross}^{1/2}$. We note that in our prior joint experiment-simulation-theory study for PnBA the experimental crosslink fraction was given by $f_{cross} = f_n/0.19$.[41] This numerical prefactor of 0.19 remains unchanged for the present $l_p/\sigma = 1.2$ model, and its value likely reflects the difference between our model construction based on the neutral pinning idea of a coarse-grained polymer model and real crosslinking of chemically complex polymers.

In polymer networks, an alternative variable to the crosslink fraction $f_n$ is the mean geometric mesh size, $a_x$. In our theory, this quantity does not play any direct role. However, based on its classic definition in experiment and simulation, $a_x$ is proportional to the square root of the average number of bare beads between two neighboring crosslinks $N_x$, i.e., $a_x \propto \sigma N_x^{1/2}$, where $f_{cross} = 1/N_x \equiv n_{crosslink}/(n_{crosslink} + n_{monomer})$. As such, we quantify this connection as $a_x = \sigma/(Af_{cross}^{\frac{1}{2}}) = \sigma/(A(f_n/0.19)^{1/2})$. Since $A/\sigma = 1/(a_x f_{cross}^{1/2})$, one can compute $A$ using experimental results for $a_x$ and $f_{cross}$. However, previous simulations of penetrant diffusion with the same coarse-grained polymer bead model [57] adopted $A = 1.94$ to achieve the best agreement of simulation data with the formula $\log(D_p) \sim (d/a_x)^2$. For this reason, we use $A = 1.94$. To check the sensitivity of our theoretical results concerning this choice, and the possible nonuniversal dependence on the precise value of $A$, we demonstrate below that using $A = 4$ does not modify any of our qualitative findings.

Based on $a_x = \sigma/A(f_n/0.19)^{1/2} \propto 1/f_n^{1/2}$ and our above discovery that $T_g(f_n) \propto f_n^{1/2}$, we can draw the theoretical deduction that $T_g(f_n)$ and $a_x^{-1}(f_n)$ are proportional. In experiment and simulation, the average mesh size $a_x$ can be directly measured [9] instead of using $a_x = \sigma/Af_{cross}^{1/2}$. Based on previous experimental results of PnBA for mesh size $a_x$ [9] and glass transition temperature $T_g(f_{cross})$ [9, 41], and our published $T_g(f_{cross})$ [41] and $a_x$ estimates from simulations [58, 72], Fig.3b tests the predicted linearity between $T_g(f_n)$ and $a_x^{-1}(f_n)$. We find excellent agreement over the entire range of data.

We now consider the ECNLE theory predictions for the polymer network alpha relaxation time for an aspect ratio of 1.2 that is modestly different than in prior work which employed 4/3 [41]. Figure 4a shows the foundational results for the nonlocal collective elastic (solid) and local cage (open) barriers as a function of $T_g(f_n)/T$ over a wide range of crosslink densities (covering the entire range achieved in experiments [9]) at various fixed temperatures. The theoretical results are plotted based on two different dynamic $T_g$ criteria appropriate for the deeply and weakly supercooled regimes in the main frame and inset of Fig.4a, respectively. Different choices of vitrification criterion do not, at zeroth order, change the qualitative finding of a good collapse and shape of the master curves. Of course, quantitatively the barriers cannot be identical, and must be larger at the same $T_g(f_n)/T$ based on defining glass formation with a longer alpha time criterion. For the classic laboratory criterion that the network vitrifies when the alpha time is 100s, the elastic barrier results as a function of $T_g(f_n)/T$ collapse very well over the *entire* $T_g/T$ range for *all* temperatures that corresponds to ~14-16 decades growth of the alpha time. The local cage barriers at different temperatures do not collapse as well, although the deviations are not large. Moreover, we find the local cage barrier grows linearly with $T_g(f_n)/T$ for each fixed temperature, while the elastic barrier grows in a non-linear manner, as discussed in great depth previously [41].

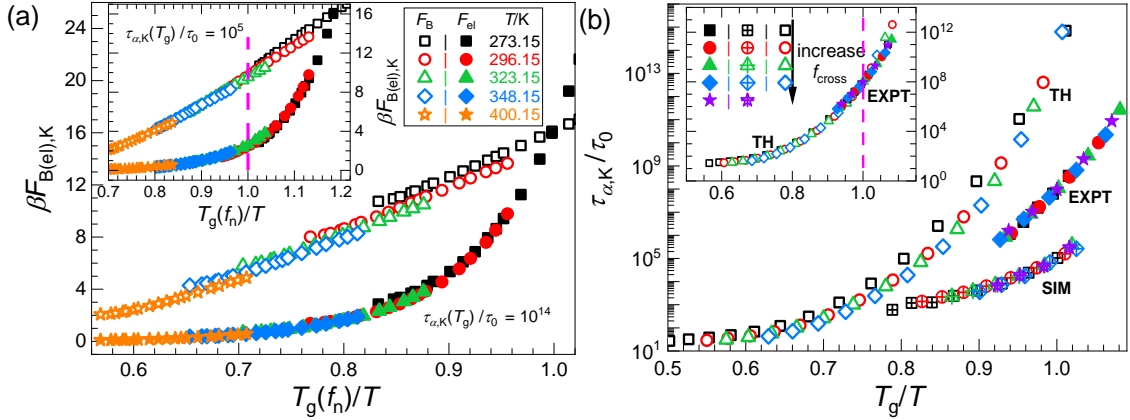

Fig.4. Kuhn segment (a) local cage and collective elastic barriers and (b) mean alpha relaxation time based on different $T_g$ criteria plotted as a function of $T_g/T$ over a wide range of crosslink densities for the $l_p/\sigma = 1.2$ model. In (b) the prior experiment and simulation results are also shown [41]. The $T_g$ criteria in the main frame and inset of (a) are $\tau_{\alpha,K}(T_g)/\tau_0 = 10^{14}$ and $10^5$ (using an elementary time scale $\tau_0 \sim 1$ps per typical viscous liquids), respectively, while for the main frame of (b) they are $\tau_{\alpha,K}(T_g) = 10^8$ps, $\tilde{\tau}_{\alpha,K}(T_g) = 10^5$ corresponding to ~320ns, and $\tau_{\alpha,K}(T_g)/\tau_0 = 10^{14}$, for experiment, simulation, and theory, respectively, and for the inset of (b) they are the same for both experiment and theory, i.e., $\tau_{\alpha,K}(T_g) = 10^8$ps or $10^8\tau_0$. Note the variable $T_g/T$ in (a) is crosslink fraction dependent at fixed temperature ($T/K = 273.15, 296.15, 323.15, 348.15, 400.15$, respectively), i.e., $T_g(f_n)$, while in (b) temperature is varied at fixed crosslink fraction ($f_n = 0, 0.02, 0.04, 0.1$, respectively). The vertical-pink-dashed lines in the insets of (a) and (b) indicate a $T_g$ criterion corresponding to a shorter relaxation time and hence larger

kinetic glass transition temperature which is well above the $T_g$ value normally deduced thermodynamically or dynamically in experiment based on a ~100 seconds timescale criterion.

Given the predicted collapse behavior of the elastic barriers and near collapse of the local cage barriers for different temperatures and crosslink densities, a nontrivial collapse of the total barrier and hence the Kuhn scale alpha times for all temperatures and crosslink densities is expected. Figure 4b verifies this is the case for the $l_\mathrm{p}/\sigma = 1.2$ model, consistent with previous work for the $l_\mathrm{p}/\sigma = 4/3$ model[41]. This behavior provides a theoretical basis for the collapsed Angell plots observed in experiment and simulation[41], as shown in the main frame of Fig.4b. Finally, we emphasize that the theoretical results in Fig.4 provide foundational input to the SCCH theory since the penetrant dynamic free energy is coupled with the polymer network dynamic free energy [34, 42, 43].

**III. SCCH Theory in Crosslinked Polymer Networks**

**A. Theoretical Background for Penetrant Dynamics**

We now briefly recall the key physical elements of the SCCH theory[34, 42, 43, 45]. This approach describes activated penetrant hopping (see the corresponding penetrant dynamic free energy in the left panel of Fig.5) which is self-consistently coupled with an activated dynamic "facilitation displacement" of polymer segments determined by the Kuhn segment dynamic free energy (see the middle panel of Fig.5). Crucially, the penetrant dynamic free energy depends on *both* the penetrant displacement *and* that of the Kuhn segment ($r_\mathrm{p}$ and $r_\mathrm{K}$, respectively). The mathematical formulation of SCCH theory has been discussed in great detail previously for a dilute penetrant in a hard sphere fluid[42, 43, 45] and in a polymer melt[34]. The derivative of the penetrant dynamic free energy in polymer melts is given as [34]:

$$\frac{\partial \beta F_{\text{dyn,p}}(r_{\text{p}},r_{\text{K}})}{\partial r_{\text{p}}} = -\frac{3}{r_{\text{p}}} + \frac{r_{\text{p}}}{3}\int \frac{d\mathbf{q}}{(2\pi)^3} q^2 \rho C_{\text{mp}}^2(q) S_{\text{mm}}(q) e^{-q^2 r_{\text{p}}^2/6} e^{-q^2 r_{\text{K}}^2 \omega_{\text{K}}(q)/6 S_{\text{mm}}(q)} \quad (5)$$

where $\rho$ is the number density of matrix interaction sites (see Fig.1), $C_{\text{mp}}(q)$ is the Fourier transform of the penetrant-polymer site-site direct correlation function, $S_{\text{mm}}(q)$ is the Fourier space static collective structure factor of the polymer network, and $\omega_{\text{K}}(q)$ is the intramolecular structure factor at the Kuhn segment level in Eq.(5) (it can be set to unity without significantly changing the numerical predictions [34]). The subscript "m" represents the elementary site level (see the left panel of Fig.1) description of the polymer that the structural correlations are determined at, which differs from the Kuhn segment scale (denoted by the subscript "K") description of dynamics. Based on the pinned bead model of polymer crosslinks[41], Eq.(5) for polymer melt is modified as:

$$\frac{\partial \beta F_{\text{dyn,p}}(r_{\text{p}},r_{\text{K}})}{\partial r_{\text{p}}} = -\frac{3}{r_{\text{p}}} + \frac{r_{\text{p}}}{3}\int \frac{d\mathbf{q}}{(2\pi)^3} q^2 C_{\text{mp}}^2(q) e^{-q^2 r_{\text{p}}^2/6}$$

$$\times [S_{\text{uu}}(q,r_{\text{Ku}}) + S_{\text{nn}}(q,r_{\text{Ku}}) + S_{\text{un}}(q,r_{\text{Ku}}) + S_{\text{nu}}(q,r_{\text{Ku}})] \quad (6)$$

The presence of pinned segments enters via the form of the collective partial Debye-Waller factors of the polymer matrix, $S_{\alpha\beta}(q,r_{\text{Ku}})$, which differs from a melt of fully mobile segments, as previously derived and discussed [41, 61]. We additionally note that the subscript notation Ku and Kn refer to the mobile (unpinned) Kuhn segments and immobile (pinned) sites, respectively.

To render SCCH theory predictive and tractable, the degree of Kuhn segment dynamic displacement that facilitates the penetrant hopping event is described by introducing a single trajectory coupling variable, $\gamma$, defined as: $r_{\text{Ku}}(\gamma) - r_{\text{loc,Ku}} = (r_{\text{p}} - r_{\text{loc,p}})/\gamma$.[34] The pinned Kuhn segments (crosslinkers) are allowed to vibrate locally with an amplitude set by the predicted localization length of unpinned Kuhn segments, i.e., $r_{\text{loc,Kn}} = r_{\text{loc,Ku}}$, per the defined notation above. The parameter $\gamma$ is determined based on enforcing a *temporal self-consistency* condition[34, [42, 43]: $\tau_{hop,p}(\gamma) = \tau_{dis,K}(\Delta r_{Ku,c}(\gamma))$, where $\Delta r_{Ku}(\gamma) = \Delta r_p(\gamma)/\gamma$ and both $\tau_{hop,p}(\gamma)$ and $\tau_{dis,K}(\Delta r_{Ku,c}(\gamma))$, are computed using Kramers theory [67, 68] as sketched in section II. Solving the self-consistency condition numerically yields $\gamma$ which depends on all control parameters of the problem. Using it, one can then compute the penetrant alpha time as $\tau_{\alpha,p} = \tau_{s,p} + \tau_{hop,p}$, where $\tau_{s,p} = d^2/D_{s,p}$ is the characteristic elementary short-time scale for the penetrant given elsewhere [34, 42]. As in the ECNLE theory of one-component liquids, at high enough matrix packing fractions the penetrant jump length can become (depending on penetrant size) sufficiently large that a collective elastic distortion of matrix particles outside the cage region is required to allow the penetrant to hop.[34, 42, 43] This physics enters via *a common* elastic barrier ($F_{el,p} = F_{el,Ku,c}$) in the computation of the timescales $\tau_{hop,p}(\gamma)$ and $\tau_{dis,K}(\Delta r_{Ku,c}(\gamma))$. The elastic barrier is determined following the same ideas and methods employed for Kuhn segment in the pure polymer system as discussed above and elsewhere, with the previously derived results given by: $\beta F_{el,p} \approx 2\pi(1 - f_n)r_{cage,p}^3 \Delta r_{eff,p}^2 \rho_K K_{0,K}$ where $\Delta r_{eff,p}(\gamma) \approx 3\Delta r_{Ku,c}^2/32 r_{cage,p}$, $r_{cage,p} = r_{min,mp}(d + N_K^{1/3}\sigma)/(d + \sigma)$, and $r_{min,mp}$ is the cage radius corresponding to the first minimum of cross radial distribution function $g_{mp}(r)$.[34, 41] Everywhere below the notation $F_{el}$ ($F_B$) is employed to indicate the penetrant elastic (local cage) barrier.

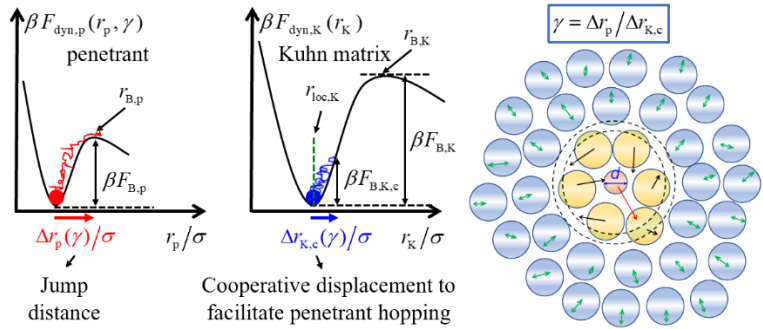

Fig.5. Schematic of the dynamic free energies for the coupled penetrant (left) and Kuhn segment (middle) displacements in SCCH theory. The trajectory level matrix facilitation of penetrant hopping is formulated based on the simplifying concept of a dimensionless dynamic coupling

parameter, $\gamma$, defined as the ratio of the penetrant jump distance $\Delta r_p(\gamma)$ to the facilitating displacement of the matrix Kuhn segments which for a non-crosslinked melt or crosslinked network is defined as $\Delta r_{K,c}(\gamma)$ or $\Delta r_{Ku,c}(\gamma)$, respectively. Relevant length and energy scales are indicated. Right panel: Schematic of the physical ideas of SCCH theory for a single smaller spherical penetrant in a bulk liquid of spheres. The image shows the matrix as disconnected spheres solely for visual clarity and is literally relevant only for the hard sphere mixture model [42, 43].

This completes our brief review of the construction of SCCH theory that includes crosslinking. From knowledge of the coupling parameter, penetrant dynamic free energy, and its local cage and elastic barriers, the SCCH theory then predicts the mean penetrant alpha time, $\tau_{\alpha,p}$, as a function of size ratio, polymer packing fraction or temperature, crosslink fraction, etc.

As discussed in the Introduction, we will apply the theory motivated by a recent experimental study of the dynamics of BTBP in PnBA networks [9]. Modeling BTBP as a sphere, its effective hard sphere diameter is 1.12nm. For the adopted $l_p/\sigma = 1.2$ Koyama model one has $\sigma$=1.23nm (for $l_p/\sigma = 4/3$ one has $\sigma$=1.03nm) thereby yielding a penetrant-to-matrix size ratio of order unity, $d/\sigma$=0.91 (1.09).

**B. Crosslink Fraction Dependence of Jump Displacements and Associated Properties**

In SCCH theory[34], the penetrant jump distance and facilitation displacement of Kuhn segments are very important for the penetrant hopping event since they (i) directly enter calculation of the penetrant elastic barrier, and (ii) the difference between displacements of a Kuhn segment *at* the penetrant and Kuhn segment alpha times provides a mechanistic picture of the degree of trajectory level coupling between penetrant and polymer. Thus, we first consider these quantities.

Both increasing crosslink fraction and decreasing temperature induces a significant increase of the penetrant and Kuhn segment alpha times. This results in an increased jump distance of both the penetrants and Kuhn segments, as shown in Figs.6a and 6b, respectively. The coupling between penetrant and Kuhn segment activated displacements at the penetrant alpha time corresponds to a facilitation displacement $\Delta r_{Ku,c}$ of a Kuhn segment, which grows with crosslink density or cooling (longer alpha time) as shown in Fig.6b. We note the crosslinking dependence of $\Delta r_{Ku,c}$ is weaker than its temperature dependence, which differs from that of $\Delta r_p$ and $\Delta r_{Ku}$ where both the temperature and crosslinking dependences are significant. We also find that the crosslink density dependence of $\Delta r_p$ is nearly temperature-independent, i.e., calculations of $\Delta r_p(f_n) - \Delta r_p(0)$ collapse (vertical shift of the data in Fig.6a) for different fixed temperatures (not shown).

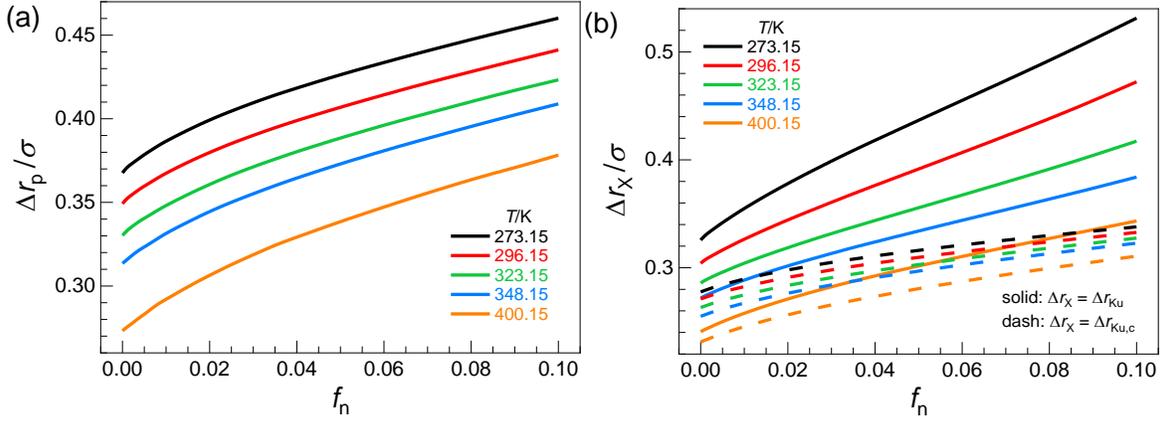

Fig.6. Jump distance of (a) the penetrant, $\Delta r_p$ and (b) unpinned mobile Kuhn segment (solid curves), $\Delta r_{Ku}$, in units of polymer bead diameter $\sigma$ at their corresponding alpha time scales as a function of crosslink fraction for a wide range of temperatures. The facilitation displacement of unpinned Kuhn segments, $\Delta r_{Ku,c}$, at the penetrant alpha time scale is shown in (b) as dashed curves. The similarity of the absolute magnitude of the penetrant and unpinned Kuhn segment jump distances reflects the strong coupling of their motions as a consequence of the very similar penetrant and Kuhn segment sizes for the model studied.

Fig.6b also shows that the absolute value of $\Delta r_{Ku,c}$ (corresponding to the penetrant alpha time scale) is much lower than that of $\Delta r_{Ku}$ (corresponding to the Kuhn segment alpha time scale), and the difference increases significantly with degree of effective supercooling (lowering temperature or increasing glass transition temperature via the crosslink fraction). This suggests that the dynamical coupling between the penetrant and polymer matrix weakens at lower temperatures and higher crosslink densities. For the highest temperature $T = 400K$ and lower crosslink density regime shown in Fig.6b, the crosslink dependence of $\Delta r_{Ku,c}$ is very similar to that of $\Delta r_{Ku}$, corresponding to a relatively stronger coupling between the alpha relaxations of the penetrant and polymer matrix.

One can also see from Fig.6b that the absolute value of the Kuhn segment facilitation displacement is not much smaller than the location of its barrier (Kuhn segment jump distance). This implies there is a strong coupling between penetrant and Kuhn segment alpha relaxations and suggests the dynamic coupling parameters, $\gamma$, defined as the ratio of penetrant jump distance $\Delta r_p(\gamma)$ to the facilitating Kuhn segment displacement $\Delta r_{Ku,c}(\gamma)$, should not be much larger than unity. Figure 7a shows this to be the case. The smaller $\gamma$ is, the larger is the dynamic coupling of activated penetrant and polymer matrix displacements. Figure 7a also shows that this coupling decreases with cooling and crosslink fraction as physically expected, but the dependences are weak. The reason is that the penetrant jump distance has a significant temperature dependence while that of Kuhn segment facilitation displacement is more limited (per Figs.6a and 6b), while both their crosslinking dependences are rather weak. Since the temporal self-consistency condition is $\tau_{hop,p}(\gamma) = \tau_{dis,K}(\Delta r_{Ku,c}(\gamma))$ with $F_{el,p} = F_{el,Ku,c}$, the coupling parameter $\gamma$ encodes dynamical displacement physics associated only with local caging. Therefore, variation of $\gamma$ with crosslink fraction is physically intuitive, and even in the large crosslink fraction regime it is nearly constant.

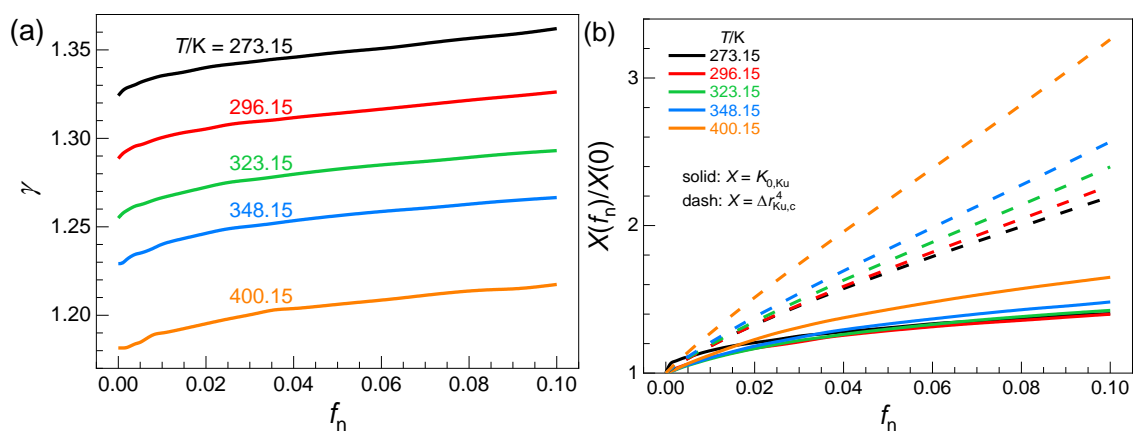

Fig.7. (a) Self-consistently determined dynamic coupling parameter, $\gamma \equiv \Delta r_p / \Delta r_{Ku,c}$, as a function of crosslink fraction at various temperatures. (b) Relative importance of the key factors $\Delta r^4_{Ku,c}$

(dash) and $K_{0,\,Ku}$ (solid) reduced by their values at zero crosslink fraction that enter the penetrant elastic barrier as a function of crosslink fraction at various temperatures.

Finally, in Fig.7b we compare the relative importance of the two key factors $\Delta r_{Ku,\,c}^4$ and $K_{0,\,Ku}$ that enter the calculation of penetrant elastic barrier. The normalized to the non-crosslinked polymer melt harmonic curvature $K_{0,\,Ku}(f_n)/K_{0,\,Ku}(0)$ ($K_{0,\,Ku}$ is proportional to the inverse dynamic localization length squared of the Kuhn segments) weakly increases with heating and crosslink fraction, with its value varying only from ~1.0-1.4 for all ranges of temperature and crosslink fractions studied. For pure hard sphere fluids, an increase of the localized state spring constant with packing fraction is the most important origin for the corresponding increase of the elastic barrier[35]. However, for networks, Fig.7b shows that as the crosslink fraction increases, the jump distance contribution $\Delta r_{Ku,\,c}^4$ dominates the change of elastic barrier. The reason is that introducing crosslinking increases the penetrant hopping time, and hence penetrants can displace significantly more corresponding to a larger jump distance, while the harmonic localization amplitude changes little.

## IV. Penetrant Alpha Relaxation Time

We now consider theoretical predictions for the penetrant alpha time. In the context of empirical experimental and simulation data[9, 56, 57], two very different choices for the crosslink fraction dependent dimensionless parameter that most faithfully encodes the effect of network crosslinking on penetrant motion can be explored. A parameter that quantifies crosslink dependent changes of segmental dynamics (activated glassy physics), or one that quantifies entropic mesh confinement. In this section we only pursue an objective empirical analysis based on plotting the theoretical data for the penetrant alpha time in two different ways. Of course, we know the physics in SCCH theory is *all* about activated hopping glassy dynamics. Our goal is to test whether a "degeneracy of interpretation" of the theoretical results might exist due to the discovered connection between $T_g$ and mesh size discussed in section II. This exercise does not answer the question whether the physics of penetrant diffusion is dominated by the effect of segmental scale activated relaxation or entropic mesh confinement, which must depend to some extent on $T_g/T$, crosslink density, and penetrant-to-matrix size ratio. This question is explored in section VI, and also in our companion simulation article[58].

As a preview of the combined insights gleaned from the results in the remainder of this article and our companion simulation paper [58], we find for the penetrant sizes studied that although crosslink fraction dependent (but temperature *independent*) mesh confinement does affect penetrant diffusivity, it is an increasingly secondary (and ultimately minor) effect as the temperature is lowered in the supercooled regime of interest compared to the coupling of penetrant hopping to the temperature and crosslink fraction dependent activated barrier hopping dynamics.

### A. Crosslinking Modified Segmental Dynamics Perspective

Figure 8 shows results for the inverse penetrant mean alpha time as a function of the temperature scaled glass transition temperature, $T_g(f_n)/T$, over a wide range of crosslink fractions at various *fixed* values of temperature. The results are presented based on a typical laboratory timescale criterion for vitrification (main panel) and a much smaller simulation-like timescale criterion (inset). The predicted behaviors are very similar for the two criteria. For each fixed temperature, the penetrant alpha time $\tau_{\alpha,p}/\tau_0$ (where $\tau_0 \sim 1$ ps per a typical viscous liquid) is predicted to increase exponentially with $T_g(f_n)$, in the spirit of an "generalized Arrhenius-like" law. Overall, results for different temperatures are rather well collapsed, as also found in our complementary simulation study [23]. There is a weak trend of lower penetrant hopping rate upon cooling, which is more prominent in the main frame that is relevant to laboratory experiments. As discussed in detail below, the reason for this difference, *and* for the roughly Arrhenius-like form of the penetrant alpha time, is that the penetrant activated relaxation time is largely controlled by the local cage barrier, not the collective elastic analog. This dominance of the local cage barrier is quantitatively even stronger in the simulation study which does not probe the deeply supercooled regime where collective elastic effects generically become increasingly important.

The results in Fig.8 can also be represented or analyzed in terms of an apparent *dimensionless* activation energy or slope of $\log(\tau_{\alpha,p}/\tau_0)$ versus $T_g(f_n)/T$ as shown in Table 1. From Table 1 and Fig.8 one sees that the dimensionless slope parameter and the absolute value of $\tau_0/\tau_{\alpha,p}$ decreases and increases with temperature, respectively. Note that if expressed in terms of an effective activation barrier in energy units, then one would have $E_A = 2.3\ k_B T_g(f_n) * slope(T)$, which is a function of crosslink fraction, $T_g$ criterion, and temperature.

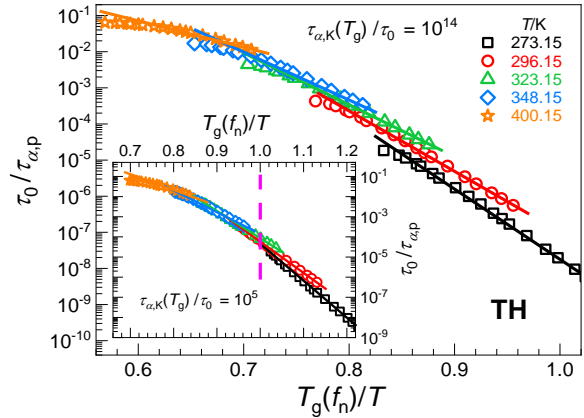

Fig.8. Penetrant inverse mean alpha time based on SCCH theory as a function of $T_g(f_n)/T$ at various fixed temperatures for vitrification criteria $\tau_{\alpha,K}(T_g)/\tau_0 = 10^{14}$ (main) and $10^5$ (inset). The vertical line in the inset is an estimate of a higher kinetic glass transition temperature, which corresponds to a simulation criterion of vitrification that delivers a glass transition temperature far higher than its laboratory-based analog.

**Table 1.** Dimensionless slope of the logarithm of the penetrant alpha time (taken to be proportional to the inverse diffusion constant) versus $T_g/T$ or $C$ plot for various fixed temperatures and $T_g$ criteria where $A=1.94$ is adopted in theoretical evaluation of $C$.

| T/K | $T_g$ criterion | 273 | 296 | 323 | 348 | 400 |
|---|---|---|---|---|---|---|
| $\log(\tau_{\alpha,p}/\tau_0)$ vs $T_g/T$ | $10^5$ | 18.4 | 15.3 | 13.5 | 11.7 | 7.3 |
| log $(\tau_{\alpha,p}/\tau_0)$ vs $T_g/T$ | $10^{14}$ | 19.4 | 17.1 | 14.8 | 12.8 | 7.5 |
| log $(1/D_p)$ vs $T_g/T$ | $10^5$ | 26.2 | 23.7 | 22.2 | 20.6 | 17.2 |
| log $(1/D_p)$ vs $T_g/T$ | $10^{14}$ | 28.0 | 25.6 | 24.2 | 22.3 | 19.2 |
| log $(\tau_{\alpha,p}/\tau_0)$ vs $C$ | | 3.3 | 2.8 | 2.2 | 1.8 | 0.77 |
| log $(1/D_p)$ vs $C$ | | 4.5 | 3.8 | 3.2 | 2.8 | 2.1 |

SCCH theory can provide the physical mechanism to understand how crosslinking modifies the penetrant alpha relaxation process by carefully examining the penetrant local cage and collective elastic barriers. Figure 9 presents theoretical results for these quantities in an Angell-like representation, where the dynamic $T_g(f_n)$ criterion appropriate for the deeply supercooled regime is adopted. For all fixed temperatures studied, the elastic barriers are far smaller than the local cage barriers. There are several reasons for this: (i) for polymer systems, as previously established, the relative importance of the elastic barrier is quantitatively modestly weaker than for spherical particle systems because of chain connectivity [37, 38], and (ii) the penetrant size is relatively small compared to the polymer Kuhn length which is known [44] to induce significant decoupling between the penetrant and matrix dynamics and a lower elastic barrier. Figure 9 also shows that the elastic barrier increases more slowly (in a practical magnitude, not functional form, sense) with $T_g(f_n)/T$ relative to that of local cage barrier. Hence, the total penetrant barrier and its growth with $T_g(f_n)/T$ is largely dominated by the local cage barrier, and for $T_g(f_n)/T \leq 1$ it is more so for the simulation timescale vitrification criterion (inset) than its laboratory analog.

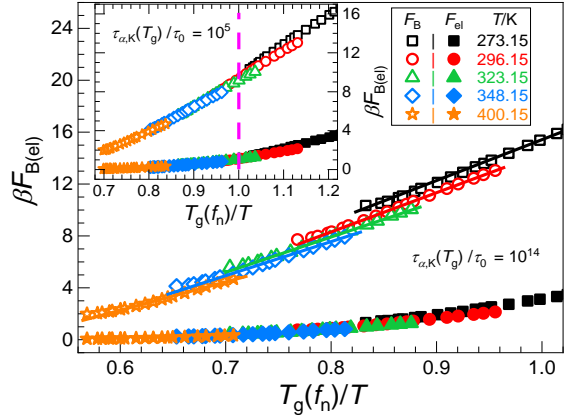

Fig.9. Penetrant local cage (open) and collective elastic (solid) barriers as a function of $T_g(f_n)/T$ at various temperatures with the $T_g$ criteria in main frame and inset corresponding to $\tau_{\alpha,K}(T_g)/\tau_0 = 10^{14}$ and $10^5$, respectively. The vertical line in the inset represents an estimate of a higher kinetic glass transition temperature corresponding to a typical simulation criterion.

More insight is obtained by comparing Figs.9 and 4a, which shows that the penetrant cage barrier is comparable to its Kuhn segment analog, as expected from the predicted strong coupling between activated penetrant hopping and Kuhn segment displacements as manifested in the close to unity value of the coupling parameter $\gamma$ in Fig.7a. On the other hand, the penetrant elastic barrier is relatively small compared to its Kuhn segment analog for the polymer network. This is because the Kuhn segment facilitation displacement is much smaller than the jump distance associated with the pure network alpha process. The latter result can be seen from Fig.6b and the fact that the jump distance enters the penetrant elastic barrier as the 4$^{th}$ power (see section III).

The local cage barrier varies linearly with $T_g(f_n)/T$ which is the same behavior predicted for the Kuhn segment (Fig.4a). The fact that this linearity is sufficiently well obeyed over the entire range of $T_g/T$, along with the smallness of the penetrant elastic barrier, is why an exponential relationship between penetrant alpha time and $T_g(f_n)/T$ is found in Fig.8. Note also from Figs.4a and 9 that at *fixed* $T_g(f_n)/T$ the cage barriers for both the penetrant and Kuhn segments are slightly larger at lower temperature. We note that the constraint of keeping fixed $T_g(f_n)/T$ implies that an increase of temperature corresponds to a corresponding compensatory increase of the crosslink fraction and hence $T_g(f_n)$. The consequences of these two changes on the penetrant alpha time tend to cancel, resulting in a near collapse of the local cage barrier (and hence relaxation rate) versus $T_g(f_n)/T$ data, to within small deviations. Overall, these considerations imply that normal glass-like physics effects play an important role in determining the temperature dependence of the penetrant alpha time, which is intensified in the high crosslink density regime. This behavior also explains why the apparent activation energies or dimensionless slopes reported in Table 1 increase as temperature decreases, a trend consistent with our simulation findings [58].

### B. Confinement Mesh Perspective

We now explore the extent to which the SCCH theory results for the penetrant alpha time can be *empirically* interpreted in a purely entropic confinement mesh scenario, $D_p \propto \exp(-BC^2)/C$ or $D_p \propto \exp(-EC)$, per Eqs (2) and (3). Specifically, we ask if the crosslink fraction dependent penetrant alpha time data can be "correlated" or "described" as: $1/\tau_{\alpha,p} \propto \exp[-(B-1)C^2]$ and/or $1/\tau_{\alpha,p} \propto \exp[-(E-b')C]$. Obviously, the entropic mesh perspective cannot account for *explicit* temperature dependences of the theoretical penetrant alpha time.

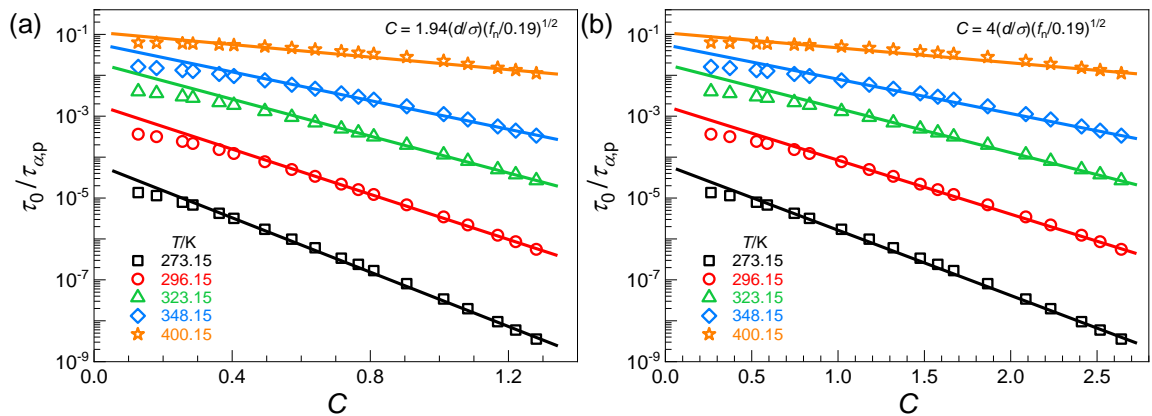

Fig.10. Inverse penetrant mean alpha time $\tau_0/\tau_{\alpha,p}$ as a function of the confinement parameter at various fixed temperatures for (a) $C = 1.94(d/\sigma)(f_n/0.19)^{1/2}$ and (b) $C = 4(d/\sigma)(f_n/0.19)^{1/2}$.

Figure 10a plots in a log-linear manner the theoretical penetrant relaxation rate as a function of confinement parameter $C = d/a_x$ over a wide range of crosslink densities and temperatures. A linear relationship is observed, albeit with a slope that decreases strongly with temperature. The latter trend is due to activated segmental scale relaxation effects, and is inconsistent with a literal [55-57] entropic confinement mesh picture (see Table 1). The theoretically predicted dependence of this slope is consistent with our simulation results in the companion paper[58]. A cautionary comment is that if simulation or experimental data is taken at only one temperature or over a narrow range, one could incorrectly conclude that the confinement mesh idea is verified. Indeed, the dimensionless slope changes documented in Table 1 prove the importance of $T$-dependent non-entropic physics in crosslinked polymer networks, and necessarily implies the rate of increase of the alpha time with $T_g/T$ is significantly larger than that with $C$. Physically the reason is simply the very different rate of change of $T_g$ and $C$ with crosslink density $f_n$. As discussed in section II, $T_g(f_n)$ is proportional to inverse mesh size $a_x^{-1}$ and hence $C(f_n)$ for a fixed penetrant size and different vitrification criteria, which is the non-causal reason the curves in Fig.10a are linearized. Our conclusions remain robust upon varying the constant $A$ in defining the confinement parameter, as shown in Fig.10b where $A = 4$ is adopted.

We also tested whether a linear relationship between $\log(\tau_{\alpha,p}/\tau_0)$ and $C^2$ can empirically "work" (not shown), i.e., $1/\tau_{\alpha,p} \sim \exp[-(B-1)C^2]$, and find it is also a reasonable description in the limited $C$ regime examined in Fig.10. However, for large enough $C$, SCCH theory predicts that the exponential relationship $1/\tau_{\alpha,p} \sim \exp[-(E - b')C]$ will break down since the penetrant elastic barrier eventually becomes non-negligible, and the total barrier does not grow linearly with confinement parameter but rather in a *stronger* quadratic manner (see Fig.9). Simulation tests of the relations $1/\tau_{\alpha,p} \propto \exp[-(B-1)C^2]$ and $1/\tau_{\alpha,p} \propto \exp[-(E-b')C]$ are reported in our companion paper, and good agreement between theory and simulation is found.[58]

**V. Decoupling Ratio of Penetrant and Network Alpha Times**

As previously studied using SCCH theory for hard sphere mixtures [42, 43] and dilute spherical penetrants in polymer melts [34], the ratio of the penetrant to Kuhn segment alpha times can be predicted which quantifies a degree of "dynamic decoupling". This quantity is especially sensitive to collective elasticity given that the hopping of penetrants and Kuhn segments generally couples to this longer-range effect to variable extent. In prior work [34, 42], the timescale decoupling ratio, $\tau_{\alpha,p}/\tau_{\alpha,K}$, was plotted as a function of matrix packing fraction and a non-monotonic behavior was predicted. Specifically, this ratio initially increases very slightly in the lower packing fraction regime corresponding to high temperature where barriers are low and the non-activated short time dynamical process timescales matter. However, with further increase of packing fraction, the timescale decoupling ratio sharply decreases indicating a strong decoupling between penetrant and matrix activated dynamics when the barriers become larger and dominate the alpha times. The crucial physical effect is the very different rate at which the penetrant and matrix elastic barriers grow with increasing packing fraction or decreasing temperature, a difference significantly larger than that for their local cage barrier analogs (see Fig.4a and Fig.9).

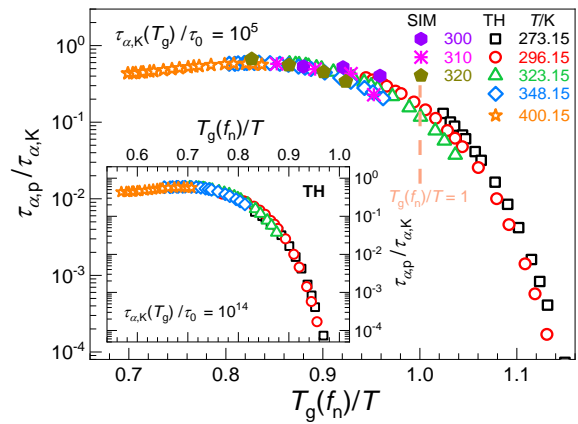

Fig.11. Decoupling time ratio as a function of $T_g(f_n)/T$ for various values of crosslink density $f_n$ using $\tau_{\alpha,K}(T_g)/\tau_0 = 10^5$ (main) and $10^{14}$ (inset) $T_g$ criteria and for five temperatures in the theory and for three temperatures for the $d/\sigma = 1.0$ simulations [58]. The simulation data of mean alpha relaxation times are based on the 1/e decay definition criterion, and the time ratio is vertically shifted up by a factor of 480 in the main panel for comparison of the functional form with the theory results. The vertical line in the main frame represents an estimate of a higher kinetic glass transition temperature based on the simulation vitrification criterion.

The inset of Figure 11 shows theoretical predictions for the decoupling time ratio plotted versus $T_g(f_n)/T$ over a wide range of $T_g(f_n)/T$ values from the rubbery regime, through the deeply supercooled regime, to the laboratory kinetic glass transition. For the four lower temperatures studied (273, 296, 323, 348 K), the time ratio decreases monotonically with the effective degree of supercooling parameter, $T_g(f_n)/T$. In contrast, at high temperature (400 K) where barriers are relatively small and only the cage barrier is non-negligible, a nearly constant (weakly increasing) behavior is predicted. Overall, for all five temperatures (which span a change of the polymer alpha time of ~14 decades), a nonmonotonic behavior is found, qualitatively consistent with the previous predictions for other systems [34, 42]. A new finding is that for all temperatures and crosslink densities, the decoupling ratio tends to collapse in $T_g(f_n)/T$ space with only slight deviations visible in the high $T_g(f_n)/T$ regime. We hope future experiments can test our results in the deeply supercooled regime where a significant decrease of the decoupling ratio is predicted.

We have verified that all the above findings remain unchanged if a different $T_g$ criterion is adopted. The main frame of Fig.11 shows an example based on a vitrification criterion relevant to simulations. For the three temperatures simulated in our companion paper [58] (300, 310, and 320 K based on the calibrated PnBA coarse-grained network model employed), the decoupling time ratio slightly decreases monotonically, and the data is in good agreement with our theoretical result in this weakly supercooled regime. We hope future simulations can better test the predicted weakly non-monotonic behavior with $T$ and probe lower temperatures.

## VI. Penetrant Diffusion Constant

### A. Introductory Comments and High-Level Picture

We now consider the predictions of Eqs. (2) and (3) which combine in a simple multiplicative manner the activated physics (at the level of mean alpha relaxation times) of SCCH theory with the entropic mesh confinement effect. We showed in section IV that the activated glassy dynamics local hopping rate predicted theoretically can be *empirically* described using the mathematical relations of the entropic mesh models formulated solely in terms of the confinement variable ( $1/\tau_{\alpha,p} \propto \exp[-(B-1)C^2]$ and $1/\tau_{\alpha,p} \propto \exp[-(E-b')C]$ ). Thus, logically, introducing the latter mechanism into Eqs. (2) or (3) [i.e., $X(C) = \exp(-C^2)/C$ or $X(C) = \exp(-b'C)$] *cannot* change the *form* of our results for how crosslink fraction modifies the diffusion constant *if* expressed in terms of $C$. Of course, the relaxation rate contribution depends on both temperature and crosslink fraction, in contrast to the mesh confinement contribution, $X(C)$.

Now, adopting Eqs.(2) or (3) for the penetrant diffusion constant, our theoretical finding that the penetrant alpha relaxation rate can be empirically described in a mesh confinement parameter framework implies it will be very difficult or impossible to dissect how important the factors $1/\tau_{\alpha,p}$ and $X(C)$ are *at fixed temperature* from experimental or simulation data. Of course, insight can be obtained by studying the temperature dependence of the penetrant relaxation time and diffusion constant. In the next subsection we consider a quantitative comparison of the theoretical predictions with and without the mesh confinement effect. In addition, as discussed in more detail in our companion simulation paper[58], reasonable agreement between SCCH theory and simulation for the crosslink fraction and temperature dependences of the penetrant diffusion constant based on Eq.(2) is found, and in a manner that *improves* with decreasing temperature. The latter trend is expected given mesh confinement is of entropic origin in contrast to the crosslinking-induced prolongation of segmental relaxation which becomes increasingly more important in colder polymer networks.

Some of the key deductions reported in our companion simulation paper [58] can also be drawn from the purely theoretical effective slope data in Table 1. One sees that under the same conditions, the effective slope determined from the diffusion constant is *larger* than that determined from the inverse alpha time, consistent with the existence of an additional mesh confinement contribution. Of course, in general, one a priori expects that the relative importance of glassy activated dynamics and mesh confinement depends on both temperature and penetrant size, with the mesh confinement (glassy activated dynamics) expected to dominate at sufficiently high (low) temperatures. Importantly, "high" and "low" depends on the specific penetrant size since the degree of coupling of penetrant activated hopping with the network segmental dynamics depends on the penetrant size relative to the size of the polymer monomer or Kuhn segment.

An additional generic physical effect in pure glass-forming liquids is "dynamic heterogeneity" (DH) which induces a distribution of activation barriers and hopping times in the lower temperature supercooled regime, and an associated decoupling of diffusivity and relaxation corresponding to a weaker temperature dependence of the self-diffusion constant compared to the alpha time [54]. The consequences of such a DH effect is not well established for penetrant dynamics, but it expected to be present at some level, and would go in the opposite direction of the entropic mesh confinement induced slowing down of diffusion relative to relaxation[58, 72]. The version of ECNLE and SCCH theories employed in this article focus solely on the mean barriers and relaxation times, and do not consider this type of DH. Thus, its absence can lead to deviations of some theoretical predictions with simulation at lower temperatures and especially for smaller penetrants due to relatively stronger effects of DH, as confirmed in our companion simulation paper.[58] On the other hand, at higher temperatures and/or for large penetrants where DH effects are either intrinsically small or are sufficiently averaged over, respectively, one expects better agreement between theory and simulation, as we have verified.[58]

We note that the ECNLE theory for the alpha relaxation in 1-component glass-forming atomic, molecular, and polymer liquids has been extended to include DH via a distribution of collective elastic barriers and relaxation times[73, 74]. It captures well the Stokes-Einstein breakdown or diffusion-relaxation decoupling and temperature-dependent stretched exponential relaxation in molecular liquids, and also the nonuniversal failure of time-temperature-superposition in polymer melts corresponding to the decoupling of the temperature dependences of the segmental alpha and chain scale end-to-end relaxation times. The ideas developed in these prior works can potentially be extended in the future to the penetrant dynamics problem in the framework of SCCH theory.

Finally, we emphasize that given the structure of Eq.(2), penetrant diffusivity will be slowed down more with crosslinking than that of penetrant alpha relaxation. This represents a qualitatively distinct form of "decoupling" of penetrant mass transport and relaxation in polymer networks compared to the Stokes-Einstein breakdown phenomenon in one-component glass forming liquids discussed above.

### B. Quantitative Theoretical Predictions

Theoretical results are presented in Fig.12 for the penetrant diffusion constant based on $D_\mathrm{p} \equiv \frac{d^2}{\tau_{\alpha,\mathrm{p}}} \exp{(-C^2)}/C$ plotted as a function of $T_\mathrm{g}(f_\mathrm{n})/T$ over a wide range of crosslink fractions for $T_\mathrm{g}$ criteria of $\tau_{\alpha,\mathrm{K}}(T_\mathrm{g})/\tau_0 = 10^{14}$ (main) and $10^5$ (inset). As found for the penetrant inverse alpha time in Fig.8, and as expected based on our insights that connect $T_g(f_n)$ and $C$ discussed in section II, we predict (i) the diffusion constant also varies exponentially with $T_g(f_n)$ with a linearity even better than that shown in Fig.8, and (ii) the slope of the logarithm of the diffusion constant versus $T_g(f_n)/T$ plot decreases with heating. However, the absolute value of diffusion constants at fixed $T_g(f_n)/T$ weakly decreases with temperature, which is opposite to our findings for inverse alpha time (see Fig.8), but in qualitative accord with our simulation findings [58]. This difference is a rather subtle 2nd order effect, but it is an objective signature of the importance of mesh confinement for the diffusion constant. Hence, we now discuss its origin in more detail.

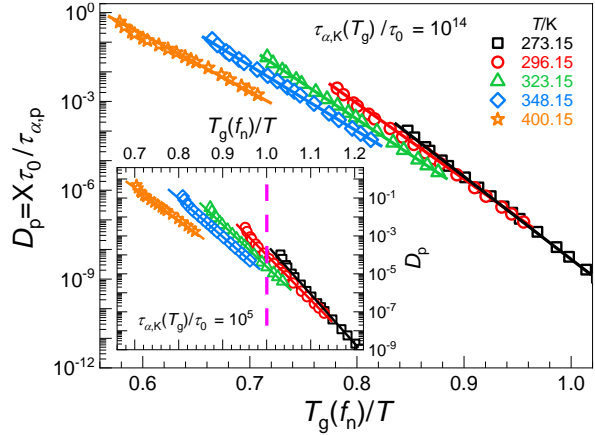

Fig.12. Penetrant diffusion constant as a function of $T_g(f_n)/T$ at various temperatures for vitrification criteria of $\tau_{\alpha,K}(T_g)/\tau_0 = 10^{14}$ (main) and $\tau_{\alpha,K}(T_g)/\tau_0 = 10^5$ (inset). The vertical line in the inset represents an estimate of a higher kinetic glass transition temperature that corresponds to a simulation-like vitrification criterion.

At fixed $T_g(f_n)/T$, a lower $T$ corresponds to a lower $T_g(f_n)$, fewer crosslinks, a larger mesh, and hence a smaller confinement parameter $C$. As discussed in the context of Figs. 8 and 9, the effect of changing temperature is slightly more important than changing $T_g(f_n)$, and thus the absolute value of the penetrant relaxation rate decreases with cooling at fixed $T_g(f_n)/T$. However, per Eq.(2), the difference between the penetrant diffusion constant and inverse alpha time is proportional to $\exp(-C^2)/C$. Thus, for the penetrant diffusion constant, the $\exp(-C^2)/C$ contribution increases the importance of crosslinking beyond that associated with the penetrant hopping rate, which, as seen in the representation of Fig.12, results in an opposite trend of the diffusion constant with cooling.

The corresponding theoretical dimensionless slope parameter results deduced from the logarithm of the diffusion constant versus $T_g(f_{cross})/T$ data is shown in Table 1. Relative to its analog for the inverse penetrant alpha time, the absolute value increases significantly to a degree that depends on temperature, $T_g$ criterion, and the chemistry-specific parameter $A$ that enters the quantification of the confinement parameter $C$. In our companion simulation paper [58], we show that by using the commonly adopted 1/$e$ decay criterion of the time correlation functions employed to define the penetrant alpha relaxation time, the degree of DH is small, and it is thus appropriate to quantitatively compare such simulation results to our theory. We find from simulation that the dimensionless slope change with temperature for $d/\sigma = 1$ is significant for the inverse alpha time (from 8.0 to 2.6), but moderate/minimal for the diffusion constant (from 9.5 to 6.1). In the present theory, the dimensionless slope change for both the inverse alpha time and diffusion constant with temperature is moderate (Table 1). We suspect this difference between theory and simulation arises from the approximate nature of Eq.(2) with $X(C) = \exp(-C^2)/C$ which reflects our simple ansatz for how to combine glassy activated physics and mesh confinement to predict diffusivity. Indeed, our simulations[58] do find modest deviations from the simple formula $D_p \sim \exp(-C^2)/C\tau_{\alpha,p}$, though they decrease with cooling as the activated segmental relaxation physics effects become relatively more important. Both theory and simulation[58] find the dimensionless slope changes with temperature for the alpha time is significantly larger than that for the diffusion constant, emphasizing the additional crosslinking effect on penetrant diffusion beyond its coupling with the activated segmental relaxation process.

As mentioned in section VIA, if $D_\mathrm{p} \sim \exp(-C^2)/C\tau_{\alpha,\mathrm{p}}$ is qualitatively correct then the diffusion constant will decrease more rapidly with $T_\mathrm{g}(f_\mathrm{n})/T$ than the penetrant alpha time increases, opposite to "normal" Stokes-Einstein relation failure[49-54] in one-component liquids associated with DH[46-48]. However, DH in pure liquids generally results in a distribution of barriers or relaxation times[73, 74], which is likely present to some degree for the problem of penetrant diffusion in glass-forming polymer networks. The pristine DH physics in pure supercooled liquids is typically manifested in a decoupling relation of the form $D_\mathrm{p}\tau_{\alpha,\mathrm{p}} \propto \tau_{\alpha,\mathrm{p}}^\delta$, where the exponent $\delta$ is modestly greater than zero. Hence, there may be two competing effects of different physical origins that can induce decoupling of penetrant diffusivity and relaxation, and which go in opposite directions with increased supercooling. As shown in our companion simulation paper [58], if the penetrant alpha time extracted from the intermediate scattering function (ISF) adopts the 1/e decay criterion to define a relaxation time, then the DH effects are relatively weak as indicated by the product $D_\mathrm{p}\tau_{\alpha,\mathrm{p}}$ *decreasing* with $T_\mathrm{g}/T$ due to the mesh confinement effect per Eq.(2) or Eq.(3). However, if the penetrant alpha time is extracted using a longer timescale criterion (the ISF decays to 0.1), then larger DH effects are present. This results in the product $D_\mathrm{p}\tau_{\alpha,\mathrm{p}}$ *increasing* with $T_\mathrm{g}/T$ corresponding to the entropic mesh confinement effect becoming of second order for the smaller penetrant studied in our simulation study [58], which is the case relevant to the present theoretical study.

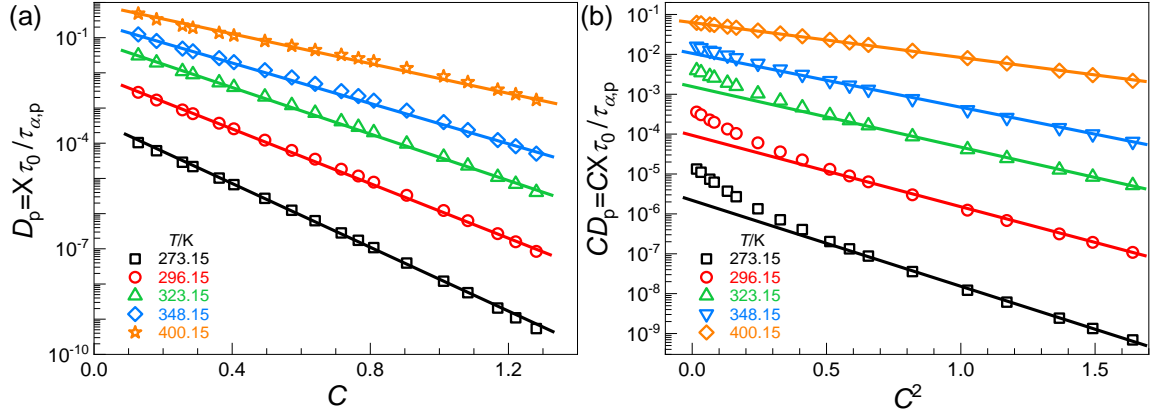

Fig.13. (a) Penetrant diffusion constant $D_p$ as a function of confinement parameter $C$ at various temperatures with $C = 1.94(d/\sigma)(f_n/0.19)^{1/2}$. (b) Same display as (a) but for the square power law relationship between $CD_p$ and $C$.

Finally, we consider our diffusion constant predictions as a function of the mesh confinement variable. Combining the exponential relation between $\log(\tau_{\alpha,p})$ and $C$ or $C^2$ predicted by SCCH theory, with an additional mesh confinement contribution per $X = \exp(-b'C)$ or $\exp(-C^2)/C$, one deduces from Eqs.(3) or (2) a net exponential relationship of the form $1/D_p \sim \exp(-EC)$ or $1/CD_p \sim \exp(-BC^2)$, respectively. Figures 13a and 13b plots the theoretical penetrant diffusion constants as a function of $C$ and $C^2$, respectively, over a wide range of temperatures and crosslink densities. One finds the diffusion constant data behaves nearly the same as that of inverse alpha time observed in Fig.10 in the following sense: (i) the diffusion constant varies exponentially with $C$ or $C^2$, and the corresponding linearity is even better (slightly worse for the $C^2$ relation relative to that based on the $C$ factor since the cage barrier dominates penetrant hopping) than that of inverse penetrant alpha time shown in Fig.10; (ii) the dimensionless slope (Table 1) of the logarithm of the inverse diffusion constant versus $C$ or $C^2$ decreases with temperature; and (iii) the diffusion constants increase with temperature. SCCH theory predicts that the elastic barrier will become important as the degree of effective supercooling grows sufficiently large (increasing penetrant size or crosslink density and/or decreasing temperature), and then both the cage and elastic barriers are crucial for penetrant hopping. In this situation, the net consequence of these two barriers results in a predicted square law relation between the *total* barrier and $C$, and hence one expects a $\log(CD_\mathrm{p}) \sim C^2$ law will emerge in the high degree of supercooling regime.

We have verified that all our conclusions above deduced from numerical computations for the diffusion constant remain qualitatively robust regardless of the adopted $T_\mathrm{g}$ criterion and choice of the prefactor *A* (1.94 is adopted in Figs.12 and 13) in defining confinement parameter *C*.

## VII. Summary and Conclusions

We have carried out a detailed theoretical study of how permanent crosslinking in polymer networks impacts the relaxation and diffusivity of dilute spherical penetrants over a wide range of temperatures, crosslink densities, and $T_\mathrm{g}$ criteria. The key new methodological aspect is a general extension of the SCCH theory of penetrant relaxation to address the dynamic consequences of chemical crosslinking based on a segment neutral pinning model. Adoption of the latter description implies the local structure remains unchanged upon crosslinking, but the microscopic dynamic constraints encountered by a penetrant from its Kuhn segment neighbors is significantly intensified as crosslink density increases. This results in a rich behavior of penetrant relaxation and diffusivity as a function of temperature, crosslink density, and penetrant size.

The penetrant local cage barrier is predicted to dominate its alpha process while the elastic barrier remains relatively small. The cage barrier varies linearly with the network glass transition temperature, leading to an exponential increase of the isothermal penetrant alpha time with $T_\mathrm{g}(f_\mathrm{n})/T$ *at* a fixed temperature. Numerical calculations based on the ECNLE theory of polymer networks reveals a proportionality between $T_\mathrm{g}$ and the square root of the crosslink fraction. We note that both $T_g$ and $f_n$ are properties of the pure polymer matrix, and hence the activated dynamics of a penetrant of fixed size is dominated by its coupling with the polymer network alpha process.

Glassy activated segmental dynamics and mesh confinement glass physics are in general both important for penetrant diffusivity, and their effects on the crosslink fraction dependence of the activation barrier are predicted to obey a remarkable degeneracy (linear in $C$ or linear in $T_g(f_n)/T$) due to the relation $C = d/a_x = A(d/\sigma)f_n^{1/2}$. However, at fixed crosslink density, the penetrant diffusivity is a strong function of temperature, a hallmark of the importance of activated segmental dynamics that is absent in the entropy-based mesh confinement effect. Overall, the latter plays an increasing minor role compared to the consequence of crosslinking induced slowing down of polymer segmental relaxation as the degree of network supercooling is increased and/or (we expect) when penetrants become sufficiently small compared to the mesh size. However, mesh confinement does slow down mass transport to varying degrees, but seemingly not the more spatially local penetrant alpha relaxation process. This results in diffusion being more strongly suppressed with increasing crosslink density than penetrant alpha relaxation, a novel form of "decoupling" of an opposite nature than the usual Stokes-Einstein breakdown[49-54] phenomenon in glass-forming liquids.

The mechanistic coupling between activated dynamics of the penetrant and network Kuhn segments at the correlated displacement level has been elucidated, and further quantified by a decoupling time ratio parameter defined as the ratio of the penetrant to Kuhn segment alpha relaxation times. This decoupling ratio varies non-monotonically as the degree of effective supercooling degree grows (increasing $T_g(f_n)/T$ via lowering temperature or increasing crosslink density), and sharply decreases in the deeply supercooled regime. The latter behavior arises because of the relative unimportance of the elastic barrier for penetrant hopping versus its high importance for Kuhn segment alpha relaxation in fragile polymer networks. A remarkably good (but not perfect) collapse of the decoupling ratio as a function of $T_g(f_n)/T$ for all crosslink densities and temperatures studied is predicted. This simplicity is found to be independent of the vitrification timescale adopted to define the glass transition temperature, and the precise value of the network strand persistence length.

Overall, we believe that the generic theoretical insights and specific predictions obtained in our work will be useful for designing new polymer networks to optimize the absolute and relative (e.g., for selectivity applications) penetrant diffusion rates relevant to polymer-based membrane separations. Extensions of the theory to treat other penetrant sizes, nonspherical penetrant shapes, inclusion of penetrant-polymer attractions, high pressure, and dynamics in the glassy state (thermosets) and vitrimers[75-78] is possible and will be studied in future work.

**Acknowledgement**

This research was supported by the U.S. Department of Energy, Office of Basic Energy Sciences, Division of Materials Sciences and Engineering (Award No. DE-SC0020858), through the Materials Research Laboratory at the University of Illinois at Urbana-Champaign. Helpful discussions with Christopher Evans are gratefully acknowledged.

**Appendix: Effect of Persistence Length Choice**

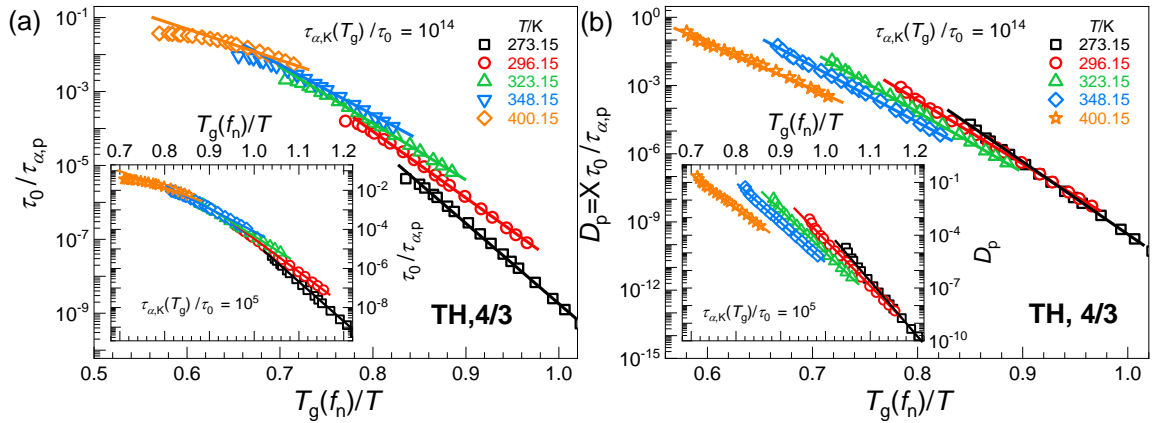

Fig.14. Penetrant (a) inverse mean alpha time and (b) diffusion constant as a function of $T_g(f_n)/T$ at various temperatures using a $\tau_{\alpha,K}(T_g)/\tau_0 = 10^{14}$ criterion. Same displays as that of the main text Fig.8 and Fig.12, respectively, but using the $l_p/\sigma = 4/3$ local aspect ratio model. Both insets are the corresponding results for a simulation-like $\tau_{\alpha,K}(T_g)/\tau_0 = 10^5$ vitrification criterion.